\documentclass[11pt,english,epsf]{scrartcl}
\usepackage{lmodern}
\usepackage[T1]{fontenc}
\usepackage[latin9]{inputenc}
\usepackage{geometry}
\usepackage{amsmath}
\usepackage{amssymb}
\usepackage{graphicx}
\usepackage{color}

\newtheorem{theorem}{Theorem}

\newcommand {\hH} {\hat{H}}
\newcommand {\hQ} {\hat{Q}}
\newcommand {\hP} {\hat{P}}
\newcommand {\hI} {\hat{I}}
\newcommand {\dfn} {\stackrel{\Delta} {=}}
\newcommand {\exe} {\stackrel{\cdot} {=}}
\newcommand {\lexe} {\stackrel{\cdot} {\le}}

\newcommand {\reals} {{\rm I\!R}}

\newcommand {\bu} {\mbox{\boldmath $u$}}
\newcommand {\bv} {\mbox{\boldmath $v$}}

\newcommand {\bx} {\mbox{\boldmath $x$}}
\newcommand {\by} {\mbox{\boldmath $y$}}

\newcommand {\bE} {\mbox{\boldmath $E$}}

\newcommand {\bU} {\mbox{\boldmath $U$}}
\newcommand {\bV} {\mbox{\boldmath $V$}}

\newcommand {\bX} {\mbox{\boldmath $X$}}
\newcommand {\bY} {\mbox{\boldmath $Y$}}

\newcommand{\calB}{{\cal B}}
\newcommand{\calC}{{\cal C}}

\newcommand{\calE}{{\cal E}}

\newcommand{\calG}{{\cal G}}

\newcommand{\calI}{{\cal I}}

\newcommand{\calT}{{\cal T}}
\newcommand{\calU}{{\cal U}}
\newcommand{\calV}{{\cal V}}

\newcommand{\calX}{{\cal X}}
\newcommand{\calY}{{\cal Y}}

\allowdisplaybreaks

\begin{document}
\thispagestyle{empty}
\title{The Generalized Stochastic Likelihood Decoder: Random Coding and
Expurgated Bounds}
\author{Neri Merhav
\thanks{This research is partially supported by the Israel Science Foundation
(ISF), grant no.\ 412/12.}
}
\date{}
\maketitle

\begin{center}
Department of Electrical Engineering \\
Technion - Israel Institute of Technology \\
Technion City, Haifa 32000, ISRAEL \\
E--mail: {\tt merhav@ee.technion.ac.il}\\
\end{center}
\vspace{1.5\baselineskip}
\setlength{\baselineskip}{1.5\baselineskip}

\begin{center}
{\bf Abstract}
\end{center}
\setlength{\baselineskip}{0.5\baselineskip}
The likelihood decoder is a stochastic decoder that selects the decoded message
at random, using the posterior distribution of the true underlying message given the
channel output. In this work, we study a generalized version of this decoder
where the posterior is proportional to a general function that depends only on
the joint empirical distribution of the output vector and the codeword. This
framework allows both mismatched versions and
universal (MMI) versions of the likelihood decoder, as well as the corresponding ordinary
deterministic decoders, among many others. We provide a direct analysis method
that yields the exact random coding exponent (as opposed to separate upper
bounds and lower bounds that turn out to be compatible, which were derived
earlier by
Scarlett {\it et al.}).
We also extend the result from pure channel coding to combined source
and channel coding (random binning followed 
by random channel coding) with side information available to the decoder.
Finally, returning to pure channel coding, we derive also an expurgated
exponent for the stochastic likelihood decoder, which turns out to be at least as tight 
(and in some cases, strictly so) as
the classical expurgated exponent of the maximum likelihood decoder, even
though the stochastic likelihood decoder is suboptimal.\\

\vspace{0.2cm}


\setlength{\baselineskip}{1.5\baselineskip}
\newpage

\section{Introduction}

The likelihood decoder for channel coding is a stochastic decoder that
selects the decoded message at random under the posterior distribution of the
correct message given the received channel output vector. The likelihood
decoder has recently received some attention, with the primary motivation
that it lends itself to considerably simpler derivations of asymptotic
upper bounds on the error probability in a variety of problems of network
information theory \cite{YAG13}. 
Owing to the duality between source encoding and channel
decoding, the likelihood encoder was also studied in the context of
rate--distortion coding \cite{SCP14}.

More recently, in \cite{SMF15} exact error exponents have been derived
for a mismatched version of the likelihood decoder, assuming a discrete memoryless
channel (DMC) and using the ensembles i.i.d.\ and 
constant composition codes. It was shown in \cite{SMF15}, among many other results,
that in the special case of the (matched) likelihood decoder, 
the random coding error exponents achieved, in both ensembles, are exactly the same as the
corresponding random coding error exponents of the optimal
maximum likelihood (ML) decoder.

The focus of this work is on further developments concerning the exact error exponent
analysis of \cite{SMF15}, as well as on extensions and refinements of this analysis in several directions.
In particular, the main contributions of this work are the following.
\begin{enumerate}
\item Allowing a more general family of stochastic likelihood decoders,
according to which the probability of deciding on a given message is
proportional to a general exponential function of the joint empirical
distribution of the codeword and the received channel output vector.
This is more general than the mismatched likelihood decoder of \cite{SMF15}.
\item Providing a direct, exponentially tight derivation of the random coding
exponent in a single analysis, instead of the separate upper and lower bounds
of \cite{SMF15} (which turn out to coincide). Hence we believe that this
analysis is somewhat simpler, at least conceptually.
\item Extending the scope to a situation of source--channel coding with side
information at the decoder, where the source coding part is based on random
binning (similarly as in \cite{jscuniv}), thus covering a variety of 
settings of theoretical and practical interest, including joint source--channel
coding with side information.
\item Returning to pure channel coding, we derive also an expurgated bound.
We point out that when this result is applied to the ordinary
likelihood decoder (which uses the real posterior probability of
each message), the resulting expurgated bound is guaranteed to be {\it at
least as tight} as the classical expurgated bound due to Csisz\'ar, K\"orner
and Marton \cite{CK11}, \cite{CKM77}, which in turn is at least as tight as
Gallager's expurgated bound \cite{Gallager68}.
This is in spite of the fact that the likelihood decoder analyzed is
suboptimal. 
We also demonstrate that the new expurgated bound may strictly improve on the
classical expurgated bound at least at high rates.
\end{enumerate}

Finally, a few comments are in order regarding the error exponent analysis. 
The analysis technique used is primarily the type class enumeration method
\cite[Chap.\ 6]{fnt}, which has already proved quite useful as a tool
for obtaining exponentially
tight random coding bounds in various contexts (see, e.g.,
\cite{REM1st}, \cite{swlist}, \cite{bid}, for a sample).
When it comes to the extension of the setup to source--channel coding with
side information, the
ensemble of codes in our setting combines random binning (for the source
coding part) and random coding (for the channel coding part), which is
somewhat more involved than ordinary error exponent analyses that involve
either one but
not both. This requires quite a careful analysis, which 
similarly as in \cite{jscuniv}, is carried out in two steps:
first, we take
the average probability of error over the ensemble of random binning codes,
for a given channel code, and at the second step, we average over the ensemble
of channel codes. 

The remaining part of the paper is organized as follows.
In Section 2, we establish notation conventions, provide some background,
and define the objectives of this paper more accurately.
In Section 3, we re-derive the exact random coding exponent of \cite{SMF15} in
an alternative way, as described above. Section 4 is devoted to the extension
to source--channel coding with side information, and finally, Section 5 is
about the expurgated bound.

\section{Notation Conventions, Background and Objectives}

\subsection{Notation Conventions}

Throughout the paper, random variables will be denoted by capital
letters, specific values they may take will be denoted by the
corresponding lower case letters, and their alphabets
will be denoted by calligraphic letters. Random
vectors and their realizations will be denoted,
respectively, by capital letters and the corresponding lower case letters,
both in the bold face font. Their alphabets will be superscripted by their
dimensions. For example, the random vector $\bX=(X_1,\ldots,X_n)$, ($n$ --
positive integer) may take a specific vector value $\bx=(x_1,\ldots,x_n)$
in $\calX^n$, the $n$--th order Cartesian power of $\calX$, which is
the alphabet of each component of this vector.
Sources and channels will be denoted by the letters $P$, $Q$, and $W$,
subscripted by the names of the relevant random variables/vectors and their
conditionings, if applicable, following the standard notation conventions,
e.g., $Q_X$, $Q_{Y|X}$, and so on. For example, the joint distribution of $(X,Y)$, induced
by $Q_X$ and $Q_{Y|X}$, will be denoted by $Q_{XY}$ and the
corresponding marginal of $Y$ will be denoted by $Q_Y$.
When there is no room for ambiguity, the
subscripts will be omitted. When we wish to refer to the joint distribution
induced by the input assignment $Q_X$ and a conditional distribution
other than $Q_{Y|X}$, say $W_{Y|X}$, we denote it by $Q_X\times W$, or simply
$Q\times W$. In this case, the marginal of $Y$, that is induced by $Q\times
W$, will be denoted by $(Q\times W)_Y$.
The probability of an event $\calE$ will be denoted by $\mbox{Pr}\{\calE\}$,
and the expectation
operator with respect to (w.r.t.) a probability distribution $Q\times P$ will be
denoted by
$\bE_{Q}\{\cdot\}$. Again, the subscript will be omitted if the underlying
probability distribution is clear from the context.
Concerning the notation of information measures,
the entropy of a random variable $X$, with a distribution $Q$, will be denoted by
$H_Q(X)$. Similarly, for a joint distribution $Q$ of $(X,Y)$, the
conditional entropy will be denoted by $H_Q(X|Y)$, the mutual
information will be denoted by $I_Q(X;Y)$, and so on. When we wish
to focus our emphasis on the dependence of the mutual information only upon the underlying
joint distribution, we denote it instead by $I(Q)$. The relative entropy (or
the Kullback--Leibler divergence) between two conditional distributions,
$Q_{Y|X}$ and $W_{Y|X}$ (or simply $W$),
weighted by the input assignment $Q_X$, will be
denoted and defined by 
\begin{equation}
D(Q_{Y|X}\|W|Q_X)=D(Q_{XY}\|Q_X\times W)=
\sum_{x\in\calX}Q(x)\sum_{y\in\calY}Q(y|x)\log\frac{Q(y|x)}{W(y|x)},
\end{equation}
where here and throughout the sequel, logarithms will be understood to be
defined with respect to the natural basis.

For two positive sequences $a_n$ and $b_n$, the notation $a_n\exe b_n$ will
stand for equality in the exponential scale, that is,
$\lim_{n\to\infty}\frac{1}{n}\log \frac{a_n}{b_n}=0$. Similarly,
$a_n\lexe b_n$ means that
$\limsup_{n\to\infty}\frac{1}{n}\log \frac{a_n}{b_n}\le 0$, and so on.
The indicator function
of an event $\calE$ will be denoted by $\calI\{E\}$. The notation $[x]_+$
will stand for $\max\{0,x\}$.

The empirical distribution of a sequence $\bx\in\calX^n$, which will be
denoted by $\hat{P}_{\bx}$, is the vector of relative frequencies
$\hat{P}_{\bx}(x)$
of each symbol $x\in\calX$ in $\bx$.
The type class of $\bx\in\calX^n$, denoted $\calT(\bx)$, is the set of all
vectors $\bx'$
with $\hat{P}_{\bx'}=\hat{P}_{\bx}$. When we wish to emphasize the
dependence of the type class on the empirical distribution $\hat{P}$, we
will denote it by
$\calT(\hat{P})$. 
Information measures associated with empirical distributions
will be denoted with `hats' and will be subscripted by the sequences from
which they are induced. For example, the entropy associated with
$\hat{P}_{\bx}$, which is the empirical entropy of $\bx$, will be denoted by
$\hat{H}_{\bx}(X)$. An alternative notation, following the conventions
described in the previous paragraph, is $H(\hP_{\bx})$.
Similar conventions will apply to the joint empirical
distribution, the joint type class, the conditional empirical distributions
and the conditional type classes associated with pairs (and multiples) of
sequences of length $n$.
Accordingly, $\hP_{\bx\by}$ would be the joint empirical
distribution of $(\bx,\by)=\{(x_i,y_i)\}_{i=1}^n$,
$\calT(\bx,\by)$ or $\calT(\hP_{\bx\by})$, will denote
the joint type class of $(\bx,\by)$, $\calT(\bx|\by)$ or
$\calT(\hP_{\bx|\by}|\by)$, will stand for
the conditional type class of $\bx$ given
$\by$, $\hH_{\bx\by}(X,Y)$ or $H(\hP_{\bx\by})$, will designate the empirical joint entropy of $\bx$
and $\by$,
$\hH_{\bx\by}(X|Y)$ will be the empirical conditional entropy,
and $\hI_{\bx\by}(X;Y)$ (or alternatively, $I(\hP_{\bx\by})$) will
denote empirical mutual information, etc. 

\subsection{Background -- The Generalized Likelihood Decoder}

Consider a DMC, designated by a matrix of single--letter input--output transition
probabilities $\{W(y|x),~x\in\calX,~y\in\calY\}$. Here the channel input
symbol $x$ takes on values in a finite input alphabet $\calX$, and the
channel output symbol $y$ takes on values in a finite output alphabet $\calY$.
When the channel is fed by a vector
$\bx=(x_1,\ldots,x_n)\in\calX^n$, it outputs a vector
$\by=(y_1,\ldots,y_n)\in\calY^n$ according to
\begin{equation}
\label{channel}
W(\by|\bx)=\prod_{t=1}^n W(y_t|x_t).
\end{equation}
A code $\calC_n\subseteq\calX^n$ is a collection of $M=e^{nR}$ channel input vectors,
$\{\bx_0,\bx_1,\ldots,\bx_{M-1}\}$, $R$ being the coding rate in nats per channel
use. It is assumed that all messages, $m=0,1,\ldots.M-1$, are equally likely.

As is very common in the information theory literature, we will consider,
throughout most of this work, the random
coding regime.
The random coding ensemble considered (here, as well as
as in \cite{SMF15}) is the ensemble of constant composition codes, where each
codeword is drawn independently under the uniform distribution within a given
type class $\calT(Q_X)$. Once the code has been randomly selected, it is revealed to
both the encoder and the decoder.

When the transmitter wishes to convey a message $m$, it transmits the
corresponding code-vector $\bx_m$ via the channel, which in turn,
stochastically maps it into an
$n$--vector $\by$ according to (\ref{channel}).
Upon receiving $\by$,
the stochastic {\it likelihood decoder} randomly selects the
estimated message $\hat{m}$ according to the induced posterior distribution
of the transmitted
message, i.e.,
\begin{equation}
\label{posterior}
\mbox{Pr}\{\hat{m}=m_0|\by\}=\mbox{Pr}\{m=m_0|\by\}=\frac{W(\by|\bx_{m_0})}{\sum_{m=0}^{M-1}W(\by|\bx_m)}.
\end{equation}
Inspired by earlier work on mismatched decoding (see, e.g., \cite{CN95},
\cite{Hui83}, \cite{MKLS94}), Scarlett {\it et al.} \cite{SMF15} studied
a mismatched version of the likelihood decoder,
which is defined similarly as in (\ref{posterior}), but with a mismatched
DMC $W'$ replacing the true one, $W$. The main results of \cite{SMF15}
are single--letter formulae for the exact random coding error exponent of the
mismatched likelihood decoder. Specifically, the random coding exponent
derived in \cite{SMF15} (see Lemma 1 therein) is given by
\begin{eqnarray}
\label{smfexp}
E(R)&=&\min_{Q_{Y|X}}\min_{\{Q_{Y|X}':~(Q_X\times Q_{Y|X}')_Y=Q_Y
\}}\left\{D(Q_{Y|X}\|W|Q_X)+\right.\nonumber\\
& &\left.[I(Q_X\times Q_{Y|X}')+[\bE_Q\log
W'(Y|X)-\bE_{Q_X\times Q_{Y|X}'}\log W'(Y|X)]_+-R]_+\right\}.
\end{eqnarray}
One of the interesting conclusions in \cite{SMF15}
is that in the special case of the regular matched likelihood decoder
($W'=W$), this expression of the
random coding error exponent coincides with that of the classical ML decoder.

\subsection{Objectives and Main Contributions}

The generalized likelihood decoder (GLD) to be considered in this work, is defined according
to
\begin{equation}
\label{gld}
\mbox{Pr}\{\hat{m}=m_0|\by\}=
\frac{\exp\{ng(\hat{P}_{\bx_{m_0}\by})\}}{\sum_{m=0}^{M-1}
\exp\{ng(\hat{P}_{\bx_m\by})\}}, 
\end{equation}
where $\hat{P}_{\bx_m\by}$ 
is the empirical distribution of $(\bx_m,\by)$ 
(whose $X$-marginal, $\hP_{\bx}$, coincides with 
$Q$) and
$g$ is a given continuous, real valued functional of this empirical distribution. 

This generalized likelihood decoder covers several important special cases.
Obviously, the choice
\begin{equation}
\label{ld}
g(\hP_{\bx_m\by})=\sum_{x,y}\hP_{\bx_m\by}(x,y)\log W(y|x) 
\end{equation}
corresponds to the ordinary
likelihood decoder. Slightly more generally, one may introduce a parameter
$\beta\ge 0$ and define
\begin{equation}
g(\hP_{\bx_m\by})=\beta\sum_{x,y}\hP_{\bx_m\by}(x,y)\log W(y|x). 
\end{equation}
Here, $\beta$ controls the degree of skewedness of the distribution
(\ref{gld}), in the spirit of the notion of finite--temperature decoding
\cite{Rujan93}: while $\beta=1$ corresponds to the usual stochastic likelihood decoder,
$\beta\to\infty$ leads to the traditional (deterministic) ML decoder. 
Likewise,
\begin{equation}
\label{mismatch}
g(\hP_{\bx_m\by})=\beta\sum_{x,y}\hP_{\bx_m\by}(x,y)\log W'(y|x) 
\end{equation}
defines a family
of mismatched likelihood decoders, bridging between the mismatched likelihood
decoder of \cite{SMF15} and the ordinary, deterministic mismatched decoder
(although the parameter $\beta$ might as well be absorbed in $W'$ in the
form of a power of $W'$). Yet another important example is
\begin{equation}
\label{mmi}
g(\hP_{\bx_m\by})=\beta I(\hP_{\bx_m\by}), 
\end{equation}
which is a parametric family of
stochastic maximum mutual information (MMI) decoders, where once again,
$\beta\to\infty$ yields the ordinary MMI universal decoder \cite{CK11}.

The main contributions in this paper are the following.
\begin{enumerate}

\item Allowing the above described more general family of stochastic
likelihood decoders (\ref{gld}) with a general function $g$. While
technically, this extension is quite straightforward,\footnote{Just replace
$\bE_Q\log W'(Y|X)$ and
$\bE_{Q_X\times Q_{Y|X}'}\log W'(Y|X)$, of (\ref{smfexp}), by $g(Q)$ and
$g(Q_X\times Q_{Y|X}')$,
respectively,} it is important to allow $g$ to be a general (not necessarily
linear) functional of the joint empirical distribution, as it covers, for example, the important
class of MMI likelihood decoders with $g$ defined as in (\ref{mmi}).

\item While in \cite{SMF15} eq.\ (\ref{smfexp}) is derived by separate
analyses of an upper bound and a matching lower bound, here we
provide directly an exponentially tight derivation of the random coding
exponent in a single analysis.
We believe that this
analysis is somewhat simpler, at least conceptually.

\item Extending the scope to a situation of source--channel coding with side
information at the decoder, where the source coding part is based on random
binning (similarly as in \cite{jscuniv}), thus covering a variety of
settings of theoretical and practical interest, including 
pure source coding, pure channel coding, joint/separate
source--channel
coding with and without side information, systematic coding, etc. Here, the
distribution of the decoded
source message given the channel output $\by$ is assumed to be
proportional to the product of two functions, the first depending on the joint
type of the source vector $\bu$ and the side information $\bv$, and the second one depends
on the corresponding code word $\bx(\bu)$ and the channel output $\by$.

\item Returning to pure channel coding, we derive also an expurgated error
exponent. An interesting point to consider is
that when this is applied to the ordinary
likelihood decoder (\ref{posterior}),
the resulting expurgated bound is guaranteed to be {\it at
least as tight} as the classical expurgated bound due to Csisz\'ar, K\"orner
and Marton \cite{CK11}, \cite{CKM77}, and
this is in spite of the fact that the likelihood decoder analyzed is
suboptimal. In this context, we study the example of the z--channel
and demonstrate that the new expurgated bound strictly improves on the
classical expurgated bound at high rates. 
\end{enumerate}

\section{Another Derivation of the Random Coding Exponent}

In this section, we provide an alternative derivation of $E(R)$, given in
(\ref{smfexp}), which is
different from the one in \cite{SMF15}, as described in item no.\ 1 above.

Assuming, without loss of generality, that message $m=0$ was
transmitted, the average probability of error of the GLD is given by
\begin{eqnarray}
\bar{P}_{\mbox{\tiny e}}&=&\bE\left\{\frac{\sum_{m=1}^{M-1}
\exp\{ng(\hat{P}_{\bX_m\bY})\}}{\sum_{m=0}^{M-1}\exp\{ng(\hat{P}_{\bX_m\bY})\}}\right\}\nonumber\\
&=&\bE\left[\bE\left\{\frac{\sum_{m=1}^{M-1}
\exp\{ng(\hat{P}_{\bX_m\bY})\}}{\sum_{m=0}^{M-1}\exp\{ng(\hat{P}_{\bX_m\bY})\}}\bigg|
\bX_0,\bY\right\}\right],
\end{eqnarray}
where the inner expectation is taken w.r.t.\ the randomness of the
incorrect codewords, $\bX_1,\ldots,\bX_{M-1}$, and the outer expectation is
taken w.r.t.\ the randomness of the transmitted codeword $\bX_0$ and the
channel output $\bY$.
We first address the inner expectation for given realizations
$(\bX_0,\bY)=(\bx_0,\by)$. Let $N_{\by}(Q')$ denote the number of codewords,
other than $\bx_0$, whose joint empirical distribution with $\by$ is given by
$Q'$. Then,
\begin{eqnarray}
\bar{P}_{\mbox{\tiny e}}(\bx_0,\by)&=&
\bE\left\{\frac{\sum_{m=1}^{M-1}
\exp\{ng(\hat{P}_{\bX_m\by})\}}{\exp\{ng(\hat{P}_{\bx_0\by})\}+
\sum_{m=1}^{M-1}\exp\{ng(\hat{P}_{\bX_m\by})\}}\right\}\nonumber\\
&=&\int_0^1\mbox{Pr}\left\{\frac{\sum_{m=1}^{M-1}
\exp\{ng(\hat{P}_{\bX_m\by})\}}{\exp\{ng(\hat{P}_{\bx_0\by})\}+
\sum_{m=1}^{M-1}\exp\{ng(\hat{P}_{\bX_m\by})\}}\ge t\right\}\mbox{d}t\nonumber\\
&=&n\cdot\int_0^\infty e^{-n\theta}\mbox{Pr}\left\{\frac{\sum_{m=1}^{M-1}
\exp\{ng(\hat{P}_{\bX_m\by})\}}{\exp\{ng(\hat{P}_{\bx_0\by})\}+
\sum_{m=1}^{M-1}\exp\{ng(\hat{P}_{\bX_m\by})\}}\ge
e^{-n\theta}\right\}\mbox{d}\theta\nonumber\\
&=&n\cdot\int_0^\infty e^{-n\theta}\mbox{Pr}\left\{(1-e^{-n\theta})\sum_{m=1}^{M-1}
\exp\{ng(\hat{P}_{\bX_m\by})\}
\ge e^{-n\theta}\exp\{ng(\hat{P}_{\bx_0\by})\}
\right\}\mbox{d}\theta\nonumber\\
&\exe&\int_0^\infty e^{-n\theta}\mbox{Pr}\left\{\sum_{m=1}^{M-1}
\exp\{ng(\hat{P}_{\bX_m\by})\}\ge
\exp\{n[g(\hat{P}_{\bx_0\by})-\theta]\}
\right\}\mbox{d}\theta\nonumber\\
&\exe&\int_0^\infty e^{-n\theta}\mbox{Pr}\left\{\sum_{Q'}N_{\by}(Q')e^{ng(Q')}
\ge\exp\{n[g(\hat{P}_{\bx_0\by})-\theta]\}
\right\}\mbox{d}\theta\nonumber\\
&\exe&\int_0^\infty e^{-n\theta}\mbox{Pr}\left\{\max_{Q'}N_{\by}(Q')e^{ng(Q')}
\ge\exp\{n[g(\hat{P}_{\bx_0\by})-\theta]\}
\right\}\mbox{d}\theta\nonumber\\
&\exe&\int_0^\infty
e^{-n\theta}\mbox{Pr}\bigcup_{Q'}\left\{N_{\by}(Q')e^{ng(Q')}
\ge\exp\{n[g(\hat{P}_{\bx_0\by})-\theta]\}
\right\}\mbox{d}\theta\nonumber\\
&\exe&\sum_{Q'}\int_0^\infty e^{-n\theta}\mbox{Pr}\left\{N_{\by}(Q')e^{ng(Q')}
\ge\exp\{n[g(\hat{P}_{\bx_0\by})-\theta]\}
\right\}\mbox{d}\theta\nonumber\\
&\exe&\max_{Q'}\int_0^\infty e^{-n\theta}\mbox{Pr}\left\{N_{\by}(Q')e^{ng(Q')}
\ge\exp\{n[g(\hat{P}_{\bx_0\by})-\theta]\}
\right\}\mbox{d}\theta\nonumber\\
&\exe&\max_{Q'}\int_0^\infty e^{-n\theta}\mbox{Pr}\left\{N_{\by}(Q')
\ge\exp\{n[g(\hat{P}_{\bx_0\by})-g(Q')-\theta]\}
\right\}\mbox{d}\theta\nonumber\\
&\dfn&\max_{Q'}\bar{P}_{\mbox{\tiny e}}(\bx_0,\by,Q'),
\end{eqnarray}
where the unions, summations and the maximizations over $\{Q'\}$ are understood to be
taken over all possible empirical distributions of sequence pairs of length
$n$, whose $X$-marginals coincide with $Q_X$.
Henceforth, for the sake of simplicity and consistency with the earlier defined
notation, we replace the notation
$\hat{P}_{\bx_0\by}$ by $Q$.
Now, given $\by$, $N_{\by}(Q')$ is a binomial random variable with $e^{nR}$ trials and success rate
of the exponential order of $e^{-nI(Q')}$. Therefore, using the techniques of
\cite[Section 6.3]{fnt}
\begin{equation}
\mbox{Pr}\left\{N_{\by}(Q')\ge\exp\{n[g(Q)-g(Q')-\theta]\}\right\}
\exe e^{-nE_1(\theta,Q,Q',R)}
\end{equation}
where
\begin{eqnarray}
E_1(\theta,Q,Q',R)&=&\left\{\begin{array}{ll}
[I(Q')-R]_+ & g(Q)-g(Q')-\theta \le [R-I(Q')]_+\\
\infty & \mbox{elsewhere}\end{array}\right.\nonumber\\
&=&\left\{\begin{array}{ll}
[I(Q')-R]_+ & \theta \ge g(Q)-g(Q')-[R-I(Q')]_+\\
\infty & \mbox{elsewhere}\end{array}\right.
\end{eqnarray}
and so,
\begin{eqnarray}
\bar{P}_{\mbox{\tiny e}}(\bx_0,\by,Q)&=&
\int_0^\infty e^{-n\theta}\mbox{Pr}\left\{N_{\by}(Q')
\ge\exp\{n[g(Q)-g(Q')-\theta]\}
\right\}\mbox{d}\theta\nonumber\\
&\exe&\int_{[g(Q)-g(Q')-[R-I(Q')]_+]_+}^\infty 
e^{-n\theta}\cdot e^{-n[I(Q')-R]_+}\mbox{d}\theta\nonumber\\
&\exe&\exp\left\{-n([I(Q')-R]_++[g(Q)-g(Q')-[R-I(Q')]_+]_+)\right\}\nonumber\\
&\dfn&e^{-nE_2(Q,Q',R)}
\end{eqnarray}
where $E_2(Q,Q',R)$ can also be written as
\begin{equation}
\label{E1}
E_2(Q,Q',R)=\left\{\begin{array}{ll}
[I(Q')- R+ g(Q)-g(Q')]_+ & R\ge I(Q')\\
I(Q')-R+[g(Q)-g(Q')]_+ & R< I(Q')
\end{array}\right.
\end{equation}
As explained briefly in \cite{SMF15},
this expression can be simplified as follows. First, for a given
$a\in\reals$ and $b\in\reals^+$, consider the identity\footnote{
To see why this identity is true, observe that if $a > b$, then $a> 0$, which means $a=[a]_+$
and the identity obviously holds. Otherwise, if $a\le b$, then $[a]_+\le b$ as
well (again, due to the positivity of $b$), in which case both $a-b$ and
$[a]_+-b$ are non--positive, and so $[a-b]_+=[[a]_+-b]_+=0$.}
$[a-b]_+=[[a]_+-b]_+$, and
applying it to the first line of (\ref{E1}) with $a=g(Q)-g(Q')$ and
$b=R-I(Q')$. Then, the first line of (\ref{E1}) can also be expressed as
$[I(Q')-R+[g(Q)-g(Q')]_+]_+$. Now, since the second line of (\ref{E1}) is
non--negative, it can also be expressed as $[I(Q')-R+[g(Q)-g(Q')]_+]_+$.
Therefore
\begin{equation}
E_2(Q,Q',R)=[I(Q')-R+[g(Q)-g(Q')]_+]_+
\end{equation}
regardless of whether $R\ge I(Q')$ or $R< I(Q')$.
Next, define
\begin{equation}
E_3(Q,R)=\min_{Q'} E_2(Q,Q',R),
\end{equation}
where the minimization is over all joint distributions $\{Q'\}$ whose $X$--marginal
is consistent with $Q_X$ and whose $Y$--marginal
agrees with $Q_Y$.
Finally, the error exponent of the GLD is given by
\begin{equation}
E(R)=\min_{Q}[D(Q\|Q_X\times W)+E_3(Q,R)],
\end{equation}
where the minimization is over all joint distributions $\{Q\}$ whose
$X$-marginal is $Q_X$. This recovers the expression (\ref{smfexp}) derived in
\cite{SMF15}.

Several comments are now in order.\\

\vspace{0.2cm}

\noindent
1. First, observe that for $g(Q)=I(Q)$, we have
\begin{equation}
E_2(Q,Q',R)=[I(Q')-R+[I(Q)-I(Q')]_+]_+=[\max\{I(Q),I(Q')\}-R]_+\ge
[I(Q)-R]_+
\end{equation}
yielding
\begin{equation}
E(R)\ge\min_{Q}\{D(Q\|Q_X\times W)+[I(Q)-R]_+\},
\end{equation}
which is exactly the random coding error exponent of the ML decoder
\cite{CK11}.
This holds true also for $g(Q)=\beta I(Q)$, provided that $\beta\ge 1$, since the
exponent is monotonically increasing in $\beta$, but on the other hand, cannot exceed the
exponent of the ML decoder. This is in analogy to the case $g(Q)=\beta\sum_{x,y}Q(x,y)\ln
W(y|x)$, which was shown in \cite{SMF15} to achieve the same exponent
as the ML decoder even for $\beta=1$, and therefore also for every $\beta\ge
1$.

\vspace{0.2cm}

\noindent
2. The highest achievable rate is calculated as follows: we seek a condition on
$R$ such that $E(R) > 0$, namely, for all $Q$ and all $Q'$ (consistent with
$Q$),
\begin{equation}
D(Q\|Q_X\times W)+[I(Q')-R+[g(Q)-g(Q')]_+]_+ > 0
\end{equation}
or, equivalently,
\begin{equation}
\max_{s,t\in[0,1]}\left\{D(Q\|Q_X\times W)+s[I(Q')-R+t[g(Q)-g(Q')]]\right\} >
0.
\end{equation}
In other words, we need that for every $Q$ and $Q'$, there exists $s$ and
$t$, both in $[0,1]$, such that
\begin{equation}
D(Q\|Q_X\times W)+s[I(Q')-R+t[g(Q)-g(Q')]] > 0.
\end{equation}
i.e., 
\begin{equation}
\forall Q,Q'~\exists~s,t\in[0,1]^2:~ R < I(Q')+t[g(Q)-g(Q')]+\frac{D(Q\|Q_X\times W)}{s}
\end{equation}
which means
\begin{eqnarray}
R&<&R_0\dfn
\min_{Q}\min_{Q'}\max_{s,t\in[0,1]}
\left\{I(Q')+t[g(Q)-g(Q')]+\frac{D(Q\|Q_X\times
W)}{s}\right\}\nonumber\\
&=&\min_{Q'}\min_Q\left\{\begin{array}{ll}
I(Q')+[g(Q_X\times W)-g(Q')]_+ & Q_{Y|X}=W\\
\infty & Q_{Y|X}\ne W\end{array}\right.\nonumber\\
&=&\min_{Q'}\{I(Q')+t[g(Q_X\times W)-g(Q')]_+\}.
\end{eqnarray}
where it should be kept in mind that the minimization is over all $\{Q'\}$ whose
$X$--marginal is $Q_X$ and whose $Y$-marginal is consistent with $Q_X\times W$.
Obviously,
\begin{equation}
R_0\le \min_Q\{I(Q)+t[g(Q_X\times W)-g(Q)]_+\}=\min\{I(Q):~g(Q)\le
g(Q_X\times W)\}\le
I(Q_X\times W).
\end{equation}
A lower on the achievable rate bound can be obtained by
\begin{equation}
R_0\ge \max_{t\in[0,1]}\min_Q\{I(Q)-tg(Q)+tg(Q_X\times W)\},
\end{equation}
which is tight when $I(Q)-tg(Q)$ is convex in $Q$ for every $t\in[0,1]$,
as is the case when $g$ is linear in $Q$ and when $g(Q)=I(Q)$.

\section{Extension to Source--Channel Coding With Side Information}

\begin{figure}[ht]
\vspace*{1cm}
\hspace*{2cm}\input{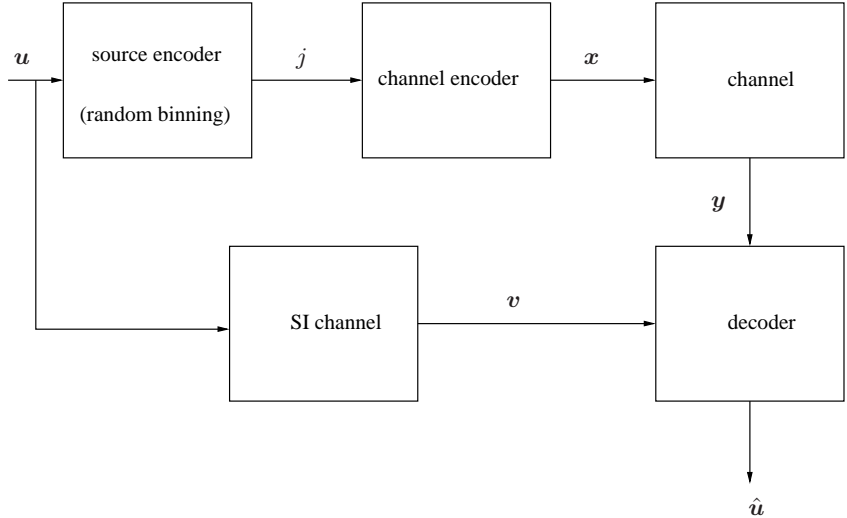}
\caption{\small Slepian--Wolf source coding, followed by channel coding. The
source $\bu$ is source--channel encoded, whereas the correlated SI $\bv$
(described as being generated by a DMC fed by $\bu$) is available at the
decoder.}
\label{sw1}
\end{figure}

Consider the communication system depicted in Fig.\ \ref{sw1}.
Let $(\bU,\bV)=\{(U_t,V_t)\}_{t=1}^n$ be $n$ independent copies of a
pair of random variables, $(U,V)\sim
P_{UV}$, taking on values in finite alphabets, $\calU$ and $\calV$,
respectively. The vector $\bU$ will designate the source vector to be encoded, whereas the
vector $\bV$ will serve as correlated side information, available to the decoder.
When a given realization $\bu=(u_1,\ldots,u_n)\in\calU^n$, of the finite alphabet source vector
$\bU$, is fed into the system, it is encoded into one out of $M=e^{nR}$
bins, selected independently at random for every member of $\calU^n$.
Here, $R>0$ is referred to as the {\it binning rate}. The bin
index $j=b(\bu)$ is mapped into a channel input vector
$\bx(j)\in\calX^n$, which in turn is transmitted across the channel $W$.
The decoder estimates $\bu$ based on the channel output $\by$ and the side
information sequence $\bv$, which is a realization of $\bV$.
As before, the various codewords $\{\bx(j)\}_{j=1}^M$ are selected independently at
random under the uniform distribution across a given type class $\calT(Q_X)$.
With a slight abuse of notation, we will sometimes denote $\bx(j)=\bx[f(\bu)]$
by $\bx[\bu]$. 

The stochastic likelihood decoder estimates $\bu$, using the channel
output $\by=(y_1,\ldots,y_n)$ and the SI vector $\bv=(v_1,\ldots,v_n)$,
according to
\begin{equation}
\mbox{Pr}\{\hat{u}=\bu_0|\bv,\by\}=
\frac{P(\bu_0,\bv)W(\by|\bx[\bu_0])}{\sum_{\bu}
P(\bu,\bv)W(\by|\bx[\bu])}.
\end{equation}
Accordingly, let us define the GLD for the source--channel coding system as
\begin{equation}
\label{jscgld}
\mbox{Pr}\{\hat{u}=\bu_0|\bv,\by\}=
\frac{\exp\{n[f(\hat{P}_{\bu_0\bv})+g(\hat{P}_{\bx(\bu_0)\by})]\}}{\sum_{\bu}
\exp\{n[f(\hat{P}_{\bu\bv})+g(\hat{P}_{\bx(\bu)\by})]\}},
\end{equation}
where $g$ is as before and similarly, $f$ is a continuous function of the
joint empirical distribution of $\bu$ and $\bv$, $\hP_{\bu\bv}$.
The average probability of error of the GLD in this setting is taken
w.r.t.\ the joint ensemble of the random binning codes and the random channel
codes described above. We refer to the asymptotic exponential rate of this
average error probability as the {\it random binning--coding error exponent}
of the GLD.

In order to characterize the random binning--coding error exponent of this GLD,
we define the following functions.
For given joint distributions $Q_{U'V}$ and $Q_{X'Y}$ of the pairs of random
variables $(U',V)$ and $(X',Y)$, respectively, we first define
\begin{equation}
h(Q_{U'V},Q_{X'Y})=f(Q_{U'V})+g(Q_{X'Y}).
\end{equation}
Next define 
\begin{equation}
E_1(R,Q_{UV})=\min_{Q_{U'V}}[[f(Q_{UV})-f(Q_{U'V})]_++R-H(U'|V)]_+,
\end{equation}
where $H(U'|V)$ is the conditional entropy of $U'$ given $V$ induced by
$Q_{U'V}$, and
\begin{equation}
E_2(R)=\min_{Q_{UV}}\{D(Q_{UV}\|P_{UV})+E_1(R,Q_{UV})\}.
\end{equation}
Now, for given joint distributions $Q_{UV}$, $Q_{XY}$, $Q_{U'V}$ and
$Q_{X'Y}$, define
\begin{equation}
E_3(Q_{UV},Q_{XY},Q_{U'V},Q_{X'Y})=[[h(Q_{UV},Q_{XY})-h(Q_{U'V},Q_{X'Y})]_++I(X';Y)-H(U'|V)]_+,
\end{equation}
where $I(X';Y)$ is the mutual information between $X'$ given $Y$ induced by
$Q_{X'Y}$, and
\begin{eqnarray}
E_4(Q_{UV},Q_{XY})&=&\min_{Q_{U'V},Q_{X'Y}}
E_3(Q_{UV},Q_{XY},Q_{U'V},Q_{X'Y}).
\end{eqnarray}
Finally, define
\begin{equation}
E_5=\min_{Q_{UV},Q_{XY}}[D(Q_{UV}\|P_{UV})+D(Q_{Y|X}\|W|Q_X)+E_4(Q_{UV},Q_{XY})].
\end{equation}
The following theorem is proved in Appendix A.
\begin{theorem}
The random binning--coding error exponent of the GLD 
(\ref{jscgld}) is given by
\begin{equation}
E(R)=\min\{E_2(R),E_5\}.
\end{equation}
\end{theorem}

\noindent
{\bf Discussion}\\ 

The term $E_2(R)$ corresponds to an error that occurs in the source
coding stage, namely, in the random
binning. It is associated with confusion of the true source vector $\bu$
with another possible source
vector $\bu'$, which is assigned to the same bin, that is, $b(\bu')=b(\bu)$.
The other term stems from the channel coding part. Here, the terms
$E_3$ and $E_4$ 
play roles that are parallel to those of $E_2$ and $E_3$ of Section 3.
In other words, every conditional type of $\{\bu'\}$ given $\bv$ can thought of
as a message set that is effectively mapped into a channel sub-code at rate $H(U'|V)$,
which is the exponential rate of the cardinality of a conditional type class.
This conditional type of source vectors competes with the true source vector
$\bu$. When the binning rate
$R$ is small, the source coding exponent $E_2(R)$ dominates, as the low binning
rate is the primary obstacle to 
reliable communication, not the channel noise. In the other extreme, when 
$R$ is very large, the binning encoder becomes a one--to--one mapping (with high
probability) and we actually pass from separate source- and channel coding to
joint source--channel coding. Consequently, the dependence on $R$ disappears.

The system considered in this section was also studied in \cite{jscuniv}, in
the context of universal decoding, with
the motivation that it provides a common umbrella to many relevant special cases,
including: separate/joint source--channel coding with/without side
information, pure source coding with decoder side information (Slepian--Wolf model),
pure channel coding, and systematic coding (see motivating discussion in
\cite{jscuniv}).
The generality of the functions $f$ and $g$ in
(\ref{jscgld}) adds considerably many additional degrees of freedom to the
model discussed, in
each of the above mentioned special cases. The various interesting choices of
$g$ have already been discussed before. Parallel choices can be considered
also for $f$, e.g., $f(Q)=\beta\bE_Q\log P'(U,V)$ for a mismatched source metric, $f(Q)=-\beta
H_Q(U|V)$ for a stochastic version of the universal minimum conditional entropy decoder,
and so on. In particular, the choice $f(Q)+g(Q')=\beta[I_{Q'}(X;Y)-H_Q(U|V)]$
is associated with a stochastic version of the universal source--channel
decoder considered in \cite{jscuniv} (which is in turn an extension of the one
in \cite{Csiszar80}). It is not difficult to verify that the universal
stochastic decoder (\ref{jscgld}), with this choice of $f(Q)+g(Q')$, achieves 
the random binning--coding error exponent of the optimal MAP decoder, for
every $\beta\ge 1$. This extends the main result of \cite{jscuniv}, which
associated with the corresponding deterministic decoder ($\beta\to\infty$).

\section{Expurgated Bound}

In this section, we return to the pure channel coding setting of Section 3 and 
derive an expurgated bound on the error probability of the GLD.
For a given code $\calC_n$, the probability of error given that message $m$ was
transmitted is given by
\begin{equation}
P_{\mbox{\tiny e}|m}(\calC_n)=\sum_{m'\ne m}\sum_{\by}W(\by|\bx_m)\cdot\frac{
\exp\{ng(\hat{P}_{\bx_{m'}\by})\}}{\exp\{ng(\hat{P}_{\bx_m\by})\}+
\sum_{m'\ne m}\exp\{ng(\hat{P}_{\bx_{m'}\by})\}}.
\end{equation}
In order to characterize the expurgated exponent, we define first a few
quantities. Let
\begin{equation}
\alpha(R,Q_Y)=\sup_{\{Q_{X|Y}:~I(Q_{XY})\le R\}}[g(Q_{XY})-I(Q_{XY})]+R,
\end{equation}
and
\begin{eqnarray}
\Gamma(Q_{XX'},R)&=&\inf_{Q_{Y|XX'}}\left\{D(Q_{Y|X}\|W|Q_X)+I_Q(X';Y|X)+\right.\nonumber\\
& &\left.[\max\{g(Q_{XY}),\alpha(R,Q_Y)\}-g(Q_{X'Y})]_+\right\}\\
&\equiv&\inf_{Q_{Y|XX'}}\left\{\bE_Q\log[1/W(Y|X)]-H(Y|X,X')+\right.\nonumber\\
& &\left.[\max\{g(Q_{XY}),\alpha(R,Q_Y)\}-g(Q_{X'Y})]_+\right\}
\end{eqnarray}

Our main result in this section is the following.
\begin{theorem}
There exists a sequence of constant composition codes, $\{\calC_n,~n=1,2,\ldots\}$,
with composition $Q_X$, such that
\begin{equation}
\liminf_{n\to\infty}\left[-\frac{\log P_{\mbox{\tiny
e}|m}(\calC_n)}{n}\right]\ge E_{\mbox{\tiny ex}}^{\mbox{\tiny gld}}(R,Q_X),
\end{equation}
where
\begin{equation}
\label{ckmstyle}
E_{\mbox{\tiny ex}}^{\mbox{\tiny gld}}(R,Q_X)=\inf_{\{Q_{XX'}:~I_Q(X;X')\le
R,~Q_{X'}=Q_X\}}[\Gamma(Q_{XX'},R)+I_Q(X;X')]-R.
\end{equation}
\end{theorem}

Note that the expression of eq.\ (\ref{ckmstyle}) has the same structure as
the Csisz\'ar--K\"orner-Marton (CKM) expurgated bound \cite{CKM77}, \cite{CK11}, except
that here the functional $\Gamma(Q_{XX'},R)$ replaces the expected Bhattacharyya
distance (under $Q_{XX'}$) that appears in the CKM expurgated bound.
The difference, however, is that unlike the expected Bhattacharyya distance, 
$\Gamma(Q_{XX'},R)$ depends, in general, on $R$. As a consequence, the
behavior of $E_{\mbox{\tiny ex}}^{\mbox{\tiny gld}}(R,Q_X)$ at high
rates is not necessarily affine as the CKM expurgated exponent,
We will return to this point later on.

\noindent
{\it Proof of Theorem 2.} Consider first the expression
\begin{equation}
Z_m(\by)\dfn\sum_{m'\ne m}\exp\{ng(\hat{P}_{\bx_{m'}\by})\}.
\end{equation}
Let $\epsilon > 0$ be arbitrary small, and for every $\by\in\calY^n$, define the set
\begin{equation}
\calB_\epsilon(m,\by)=\left\{\calC_n:~Z_m(\by)\le\exp\{n\alpha(R-\epsilon,\hat{P}_{\by})\}\right\}.
\end{equation}
In Appendix B, we show that the vast majority of constant composition codes $\{\calC_n\}$
(whose composition is $Q_X$),
are outside $\calB_\epsilon(m,\by)$, simultaneously for all $m$ and all $\by$.
More precisely, it is shown in Appendix B that, 
considering the ensemble of randomly selected constant codes
of type $Q_X$, 
\begin{equation}
\label{appb}
\mbox{Pr}\{\calB_\epsilon(m,\by)\}\le \exp\{-e^{n\epsilon}+n\epsilon+1\},
\end{equation}
for every $m$ and $\by$, and so, by the union bound, this means that
\begin{equation}
\mbox{Pr}\left\{\bigcup_{m}\bigcup_{\by\in\calY^n}\calB_\epsilon(m,\by)\right\}\dfn
\mbox{Pr}\{\calB_\epsilon\}\le
e^{nR}|\calY|^n\exp\{-e^{n\epsilon}+n\epsilon+1\},
\end{equation}
which still decays double--exponentially.
Thus, for all codes in $\calG_\epsilon=\calB_\epsilon^c$, which is 
the vast majority of constant composition codes codes $\{\calC_n\}$ with
composition $Q_X$, we have $Z_m(\by)\ge
\exp\{n\alpha(R-\epsilon,\hat{Q}_{\by})\}$ simultaneously for all
$m=0,1,\ldots,M-1$ and $\by\in\calY^n$.
Now, trivially,
\begin{equation}
\frac{\exp\{ng(\hat{P}_{\bx_{m'}\by})\}}{\exp\{ng(\hat{P}_{\bx_m\by})\}+
\sum_{m'\ne m}\exp\{ng(\hat{P}_{\bx_{m'}\by})\}} \le 1,
\end{equation}
and for a code in $\calG_\epsilon\dfn\calB_\epsilon^c$, we also have
\begin{equation}
\frac{\exp\{ng(\hat{P}_{\bx_{m'}\by})\}}{\exp\{ng(\hat{P}_{\bx_m\by})\}+
\sum_{m'\ne m}\exp\{ng(\hat{Q}_{\bx_{m'}\by})\}} \le 
\frac{\exp\{ng(\hat{P}_{\bx_{m'}\by})\}}{\exp\{ng(\hat{P}_{\bx_m\by})\}+
\exp\{n\alpha(R-\epsilon,\hat{P}_{\by})\}}.
\end{equation}
Thus, for such a code
\begin{equation}
\frac{\exp\{ng(\hat{P}_{\bx_{m'}\by})\}}{\exp\{ng(\hat{P}_{\bx_m\by})\}+
\sum_{m'\ne m}\exp\{ng(\hat{P}_{\bx_{m'}\by})\}} \le \min\left\{1,
\frac{\exp\{ng(\hat{P}_{\bx_{m'}\by})\}}{\exp\{ng(\hat{P}_{\bx_m\by})\}+
\exp\{n\alpha(R-\epsilon,\hat{P}_{\by})\}}\right\}.
\end{equation}
It follows that for every $\calC_n\in\calG_\epsilon$,
\begin{eqnarray}
P_{\mbox{\tiny e}|m}(\calC_n)&\le&\sum_{m'\ne
m}\sum_{\by}W(\by|\bx_m)\cdot\min\left\{1,\frac{
\exp\{ng(\hat{P}_{\bx_{m'}\by})\}}{\exp\{ng(\hat{P}_{\bx_m\by})\}+
\exp\{n\alpha(R-\epsilon,\hat{P}_{\by})\}}\right\}\nonumber\\
&\exe&\sum_{m'\ne m}\sum_{\by}W(\by|\bx_m)
\exp\{-n[\max\{g(\hP_{\bx_m\by}),
\alpha(R-\epsilon,\hat{P}_{\by})\}-g(\hat{P}_{\bx_{m'}\by})]_+\}\nonumber\\
&\exe&\sum_{m'\ne
m}\exp\{-n\Gamma(\hat{P}_{\bx_m\bx_{m'}},R-\epsilon)\}\nonumber\\
&=&\sum_{Q_{X'|X}:~Q_{X'}=Q_X}N_m(Q_{XX'})\exp\{-n\Gamma(Q_{XX'},R-\epsilon)\}
\end{eqnarray}
where $N_m(Q_{XX'})$ is the number of codewords $\{\bx_{m'}\}$ whose joint
type with $\bx_m$ is exactly $Q_{XX'}$,
and where 
\begin{eqnarray}
\label{summingup}
& & \sum_{\by}W(\by|\bx_m)
\exp\{-n[\max\{g(\hat{Q}_{\bx_m\by}),\alpha(R-\epsilon,\hat{Q}_{\by})\}
-g(\hat{Q}_{\bx_{m'}\by})]_+\}\nonumber\\
&\exe&
\max_{Q_{Y|XX'}}\exp\{n(H_Q(Y|X,X')-H_Q(Y|X)-D(\hQ_{Y|X}\|W|P)-\nonumber\\
& &[\max\{g(Q_{XY}),\alpha(R-\epsilon,Q_Y)\}-g(Q_{X'Y})]_+)\}
\nonumber\\
&=&\exp\left\{-n\min_{Q_{Y|XX'}}\left[D(\hQ_{Y|X}\|W|P)+I_Q(X';Y|X)+\right.\right.\nonumber\\
& &\left.\left.[\max\{g(Q_{XY}),\alpha(R-\epsilon,Q_Y)\}-g(Q_{X'Y})]_+\right]\right\}
\nonumber\\
&=& \exp\{-n\Gamma(Q_{XX'},R-\epsilon)\}.
\end{eqnarray}
Now, as is shown in Appendix C, for most codes in $\calG_\epsilon$,
\begin{equation}
\label{appc}
N_m(Q_{XX'})\le \left\{\begin{array}{ll}
\exp\{n[R-I_Q(X;X')]\} & R\ge I_Q(X;X')\\
0 & R< I_Q(X;X')\end{array}\right.
\end{equation}
for all $m$ and all $Q_{XX'}$, and so, considering the arbitrariness of
$\epsilon$, the expurgated error exponent is given by
\begin{equation}
E_{\mbox{\tiny ex}}^{\mbox{\tiny gld}}(R,Q_X)=
\min_{\{Q_{XX'}:~I_Q(X;X')\le, Q_{X'}=Q_X\}}[\Gamma(Q_{XX'},R)+I_Q(X;X')]-R
\end{equation}
This completes the proof of Theorem 2.
$\Box$

It is interesting to note an important difference between the first steps in
the derivation in the proof of Theorem 2 above, and the first steps in the
derivation of the ordinary expurgated bound.
While for the ordinary expurgated bound, the starting point is the inequality
\begin{equation}
P_{\mbox{\tiny e}|m}(\calC_n)=\sum_{m'\ne m}\sum_{\by}W(\by|\bx_m)\cdot
\sqrt{\frac{W(\by|\bx_{m'})}{W(\by|\bx_m)}}
\end{equation}
or, more generally,
\begin{equation}
P_{\mbox{\tiny e}|m}(\calC_n)=\sum_{m'\ne m}\sum_{\by}W(\by|\bx_m)\cdot
\left[\frac{e^{ng(\hP_{\bx_{m'}\by})}}{e^{ng(\hP_{\bx_m\by})}}\right]^\gamma,~~~~
\gamma\ge 0,
\end{equation}
the above derivation in the proof of Theorem 2 begins from
from the inequality
\begin{equation}
P_{\mbox{\tiny e}|m}(\calC_n)=\sum_{m'\ne m}\sum_{\by}W(\by|\bx_m)\cdot
\min\left\{1,\frac{\exp\{ng(\hat{P}_{\bx_{m'}\by})\}}{\exp\{ng(\hat{P}_{\bx_m\by})\}+
\exp\{n\alpha(R-\epsilon,\hat{P}_{\by})\}}\right\}.
\end{equation}
It is easy to argue that for $\gamma\in[0,1]$ (and in particular,
$\gamma=1/2$, used at least when $g(Q)=\sum_{x,y}Q(x,y)\ln W(y|x)$):
\begin{equation}
\label{comp}
\min\left\{1,\frac{\exp\{ng(\hat{P}_{\bx_{m'}\by})\}}{\exp\{ng(\hat{P}_{\bx_m\by})\}+
\exp\{n\alpha(R-\epsilon,\hat{P}_{\by})\}}\right\}\le
\left[\frac{e^{ng(\hP_{\bx_{m'}\by})}}{e^{ng(\hP_{\bx_m\by})}}\right]^\gamma.
\end{equation}
To see why this is true, let us distinguish between the cases
$g(\hat{P}_{\bx_{m'}\by})\le g(\hat{P}_{\bx_m\by})$ and
$g(\hat{P}_{\bx_{m'}\by})> g(\hat{P}_{\bx_m\by})$.
In the former case,
\begin{eqnarray}
& &\min\left\{1,\frac{\exp\{ng(\hat{P}_{\bx_{m'}\by})\}}{\exp\{ng(\hat{P}_{\bx_m\by})\}+
\exp\{n\alpha(R-\epsilon,\hat{P}_{\by})\}}\right\}\\
&\le&
\frac{\exp\{ng(\hat{P}_{\bx_{m'}\by})\}}{\exp\{ng(\hat{P}_{\bx_m\by})\}}\\
&\le&\left[\frac{\exp\{ng(\hat{P}_{\bx_{m'}\by})\}}{\exp\{ng(\hat{P}_{\bx_m\by})\}}\right]^\gamma.
\end{eqnarray}
In the latter case, the right--hand side of (\ref{comp}) exceeds unity,
whereas the left--hand
side is always less than unity. Since all the subsequent derivations in the
proof of Theorem 2 are exponentially tight (by the method of types),
the conclusion from this observation is that at least for the choice
$g(Q)=\sum_{x,y}Q(x,y)\log W(y|x)$,
the new expurgated bound, $E_{\mbox{\tiny ex}}^{\mbox{\tiny gld}}(R,Q_X)$, is {\it at least as
tight} as the CKM expurgated bound. In the next example, we demonstrate that
it may indeed be strictly tighter than the CKM expurgated bound at least at
relatively high rates.\\

\begin{figure}[h!t!b!]
\centering
\includegraphics[width=8.5cm, height=8.5cm]{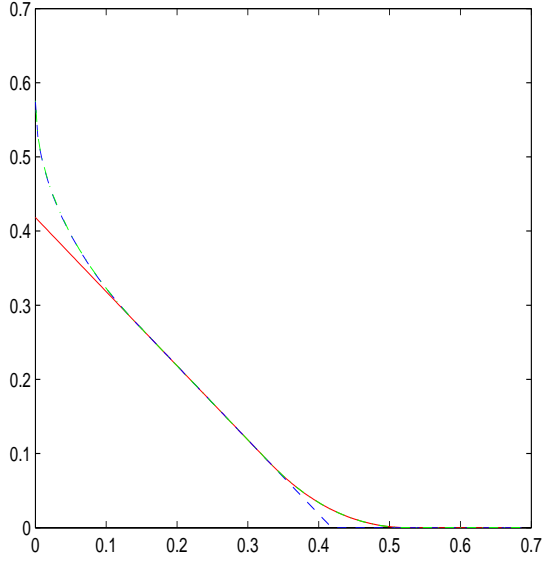}
\caption{Various exponents for the z--channel with parameter $w=0.9$.
The red (solid) curve is the random coding bound, the blue (dashed)
one is the classical
expurgated bound, and the green (dashed)
curve is the new expurgated bound. The latter
seems to behave as the maximum between the first two.}
\label{bounds}
\end{figure}

\noindent
{\it Example -- the Z-Channel.}
Consider the z--channel with $\calX=\calY=\{0,1\}$, which is parametrized by $w\in[0,1]$ as follows:
\begin{equation}
W(y|x)=\left\{\begin{array}{ll}
w & x=y=0\\
1-w & x=0,~y=1\\
0 & x=1,~y=0\\
1 & x=y=1\end{array}\right.
\end{equation}
and let the input assignment be $Q_X(0)=Q_X(1)=1/2$.
Let $g(Q)=\bE_Q\log W(Y|X)$. In the case of the z--channel,
any joint empirical distribution $Q_{XY}$ for which $g(Q_{XY}) > -\infty$,
must also be of the z--form:
\begin{equation}
Q_{XY}(x,y)=\left\{\begin{array}{ll}
q/2 & x=y=0\\
(1-q)/2 & x=0,~y=1\\
0 & x=1,~y=0\\
1/2 & x=y=1\end{array}\right.
\end{equation}
where $q\in[0,1]$ designates the associated empirical transition probability
from $X=0$ to $Y=0$.
Thus,
\begin{equation}
Q_Y(y)=\left\{\begin{array}{ll}
q/2 & y=0\\
1-q/2 & y=1\end{array}\right.
\end{equation}
Now,
\begin{equation}
g(Q_{XY})=g(q)\dfn\left\{\begin{array}{ll}
\frac{q}{2}\log w+\frac{1-q}{2}\log(1-w) & Q_{XY}(0,1)=0\\
-\infty & Q_{XY}(0,1)>0\end{array}\right.
\end{equation}
We begin from the calculation of $\alpha(R,Q_Y)$, which will be denoted by
$\alpha(R,q)$.
We observe that for a given $Q_Y$, which means actually, a given $q$, there is
only one empirical channel, so here, the set $\{Q_{X|Y}: I(Q_{XY})\le R\}$ is
either a singleton or an empty set, depending on $q$ and $R$.
The mutual information for a given $q$ is
\begin{equation}
I(Q_{XY})=I(q)=h\left(\frac{q}{2}\right)-\frac{1}{2}h(q),
\end{equation}
where $h(\cdot)$ is the binary entropy function.
Thus,
\begin{equation}
\alpha(R,q)=\left\{\begin{array}{ll}
\frac{q}{2}\log w+\frac{1-q}{2}\log(1-w)-I(q)+R & I(q)\le R\\
-\infty & I(q) > R\end{array}\right.
\end{equation}
and so,
\begin{equation}
\max\{g(Q_{XY}),\alpha(R,Q_Y)\}=\max\{g(q),\alpha(R,q)\}=g(q)+[R-I(q)]_+.
\end{equation}
which yields
\begin{equation}
\left[\max\{g(Q_{XY}),\alpha(R,Q_Y)\}-g(Q_{X'Y})\right]_+=[g(q)+[R-I(q)]_+-g(q)]_+=[R-I(q)]_+.
\end{equation}
For a given $q$, which is actually a given $\hat{P}_{\by}$, and a given pair of
codewords $\{\bx_m,\bx_{m'}\}$, with a joint empirical distribution
$\hQ_{XX'}$. we are summing in (\ref{summingup}) the expression
$e^{ng(q)}\cdot e^{-n[R-I(q)]_+}$ over all $\by$, but the summand is positive only
for $\by$ for which
both $\hP_{\bx_m\by}$ and
$\hP_{\bx_{m'}\by}$ agree with $Q_{XY}$ as defined above (with $q$).
This can be the case only if $q \le 2Q_{XX'}(0,0)$ and
\begin{eqnarray}
Q_{Y|XX'}(0|0,0)&=&
\frac{q}{2Q_{XX'}(0,0)}\\
Q_{Y|XX'}(1|0,0)&=&
1-\frac{q}{2Q_{XX'}(0,0)}\\
Q_{Y|XX'}(0|0,1)&=&
Q_{Y|XX'}(0|1,0)=
Q_{Y|XX'}(0|1,1)=0.
\end{eqnarray}
The above--mentioned sum is therefore of the exponential order of
$$\exp\left\{n\left[Q_{XX'}(0,0)h_2\left(\frac{q}{2Q_{XX'}(0,0)}\right)
+g(q)-[R-I(q)]_+\right]\right\}.$$
and so,
\begin{equation}
\Gamma(Q_{XX'},R)=[R-I(q)]_+-g(q)-Q_{XX'}(0,0)\cdot
h\left(\frac{q}{2Q_{XX'}(0,0)}\right).
\end{equation}
Let us denote $\theta=Q_{XX'}(0,0)$ ($\theta\le 1/2$), so
\begin{equation}
\Gamma(\theta)=[R-I(q)]_+-g(q)-\theta h\left(\frac{q}{2\theta}\right).
\end{equation}
Note that since both marginals of $Q_{XX'}$ are binary symmetric sources, then
$Q_{XX'}(1,0)=
Q_{XX'}(0,1)=1/2-\theta$ and
$Q_{XX'}(1,1)=\theta$.
Now,
\begin{equation}
I(Q_{XX'})=I(\theta)\dfn
2\theta\log\frac{\theta}{1/4}+2\left(\frac{1}{2}-\theta\right)\log
\frac{1/2-\theta}{1/4}=\log 2-h(2\theta).
\end{equation}
It follows that
$Q_{XX'}(1,0)=
Q_{XX'}(0,1)=1/2-\theta$ and
$Q_{XX'}(1,1)=\theta$.
Now,
\begin{equation}
I(Q_{XX'})=I(\theta)\dfn
2\theta\log\frac{\theta}{1/4}+2\left(\frac{1}{2}-\theta\right)\log
\frac{1/2-\theta}{1/4}=\log 2-h(2\theta).
\end{equation}
It follows that
\begin{equation}
\label{zchannel}
E_{\mbox{\tiny ex}}^{\mbox{\tiny gld}}(R,Q_X)=
\min_{\{\theta:~\log 2-h(2\theta)\le R,~\theta\le 1/2\}}\min_{q\le
2\theta}\left\{
[R-I(q)]_+-g(q)-h(2\theta)-\theta
h\left(\frac{q}{2\theta}\right)\right\}+\log 2-R.
\end{equation}
The ordinary expurgated bound (CKM), on the other hand, is given by
\begin{equation}
\label{ckmz}
E_{\mbox{\tiny ex}}(R,Q_X)=\left\{\begin{array}{ll}
-\frac{1}{2}h^{-1}(\log 2-R)\log(1-w) &
R\le \log 2- h\left(\frac{1}{1+\sqrt{1-w}}\right)\\
\log\frac{2}{1+\sqrt{1-w}}-R & R > \log 2-h\left(\frac{1}{1+\sqrt{1-w}}\right)
\end{array}\right.
\end{equation}
Interestingly, if the non--negative term
$[R-I(q)]_+$ is discarded from
(\ref{zchannel}), which results in a lower bound to
$E_{\mbox{\tiny ex}}^{\mbox{\tiny gld}}(R,Q_X)$,
then the minimization of the remaining expression can easily be carried out
analytically, and it turns out to yield exactly 
the same expression as that of the CKM expurgated exponent in eq.\ (\ref{ckmz}).
Thus, it is the term $[R-I(q)]_+$ that has the potential to improve on the CKM
expurgated exponent, at least at relatively high rates. Indeed, 
in Fig.\ \ref{bounds}, we see comparative plots of the CKM expurgated exponent
(in blue) and the new expurgated exponent (in green) as well as the random
coding error exponent (in red), all for $w=0.9$. As can be seen, while the CKM expurgated
exponent descends linearly for high rates (as is well known), the new expurgated exponent
departs from it in the high rate region, and it seems to follow the curve of
the random coding exponent.
In other words, at least in this example, the new expurgated exponent seems to
follow the maximum between the random coding exponent and the CKM exponent,
and therefore to improve on the CKM expurgated exponent at high rates.
This is in spite of the fact that the new expurgated exponent was developed
for a sub-optimal decoder. We believe that one of the reasons for this
improvement is that the new expurgated bound is not based on the union bound,
which is inherently the starting point of the classic expurgated bound, and
also its weakness at high rates.

In future research, it would be interesting
to explore the new expurgated bound in additional examples 
and see if it may improve on existing lower
bounds to the reliability function, and thereby shrink the gap between the
well known lower bounds and upper bounds to this function.

\section*{Appendix A}
\renewcommand{\theequation}{A.\arabic{equation}}
    \setcounter{equation}{0}

\noindent
{\it Proof of Theorem 1.}
The proof is based on the same technique as in the derivation in Section 3,
as well as in \cite{jscuniv}.
The probability of error is given by
\begin{equation}
\bar{P}_e=\bE\left\{\frac{\sum_{\bu'\ne\bU}
\exp\{n[f(\hat{Q}_{\bu'\bV})+g(\hat{Q}_{\bX(\bu')\bY})]\}}
{\sum_{\bu'}\exp\{n[f(\hat{Q}_{\bu'\bV})+g(\hat{Q}_{\bX(\bu')\bY})]\}}\right\}.
\end{equation}
Let us condition first on
$(\bU=\bu_0,\bV=\bv,b(\bu_0)=j_0,\bX(j_0)=\bx_0,\bY=\by)$ and
take the expectation only w.r.t.\ the random binning of source vectors
other than $\bu_0$ and codewords other than $\bX(j_0)$.
Using the same technique as before, we assess the conditional probability of
error as
\begin{eqnarray}
& &\bar{P}_{\mbox{\tiny e}}(\bu_0,\bv,j_0,\bx_0,\by)\nonumber\\
&\exe&\int_0^\infty e^{-n\theta}\cdot\mbox{Pr}\left\{
\sum_{\bu'}\exp\{n[f(\hat{Q}_{\bu'\bv})+g(\hat{Q}_{\bX(\bu')\by})]\}
\ge\right.\nonumber\\
&
&\left.\exp\{n[f(\hat{Q}_{\bu_0\bv})+g(\hat{Q}_{\bX(\bu_0)\by})-\theta]\}\right\}\mbox{d}\theta.
\end{eqnarray}
We first condition on the channel code
$\calC_n=\{\bx_0,\bx_1,\ldots,\bx_{M-1}\}$,
$M=e^{nR}$, and calculate the probability only w.r.t.\ the randomness of the binning.
Consider the following decomposition:
\begin{eqnarray}
& &\sum_{\bu'}\exp\{n[f(\hat{P}_{\bu'\bv})+g(\hat{P}_{\bx(\bu')\by})]\}\nonumber\\
&=&
\sum_{\bu':b(\bu')=b(\bu)}\exp\{n[f(\hat{P}_{\bu'\bv})+g(\hat{P}_{\bx(\bu')\by})]\}+
\sum_{\bu':b(\bu')\ne
b(\bu)}\exp\{n[f(\hat{P}_{\bu'\bv})+g(\hat{P}_{\bx(\bu')\by})]\}\nonumber\\
&=&\exp\{ng(\hat{P}_{\bx(\bu)\by})\}\cdot
\sum_{\bu':b(\bu')=b(\bu)}\exp\{nf(\hat{P}_{\bu'\bv})\}+
\sum_{\bu':b(\bu')\ne
b(\bu)}\exp\{n[f(\hat{P}_{\bu'\bv})+g(\hat{P}_{\bx(\bu')\by})]\}\nonumber\\
&\dfn& Z_1+Z_2.
\end{eqnarray}
Then, obviously,
\begin{eqnarray}
& &\mbox{Pr}\left\{
\sum_{\bu'}\exp\{n[f(\hat{P}_{\bu'\bv})+g(\hat{P}_{\bx(\bu')\by})]\}
\ge
\exp\{n[f(\hat{P}_{\bu_0\bv})+g(\hat{P}_{\bx_0\by})-\theta]\}\right\}\nonumber\\
&\exe&\mbox{Pr}\left\{Z_1\ge
\exp\{n[f(\hat{P}_{\bu_0\bv})+g(\hat{P}_{\bx_0\by})-\theta]\}\right\}+\nonumber\\
& &\mbox{Pr}\left\{Z_2\ge
\exp\{n[f(\hat{P}_{\bu_0\bv})+g(\hat{P}_{\bx_0\by})-\theta]\}\right\}.
\end{eqnarray}
Let us begin with the first term,
\begin{eqnarray}
& &\mbox{Pr}\left\{Z_1\ge
\exp\{n[f(\hat{P}_{\bu\bv})+g(\hat{P}_{\bx(\bu)\by})-\theta]\}\right\}\nonumber\\
&=&\mbox{Pr}\left\{\sum_{\bu':b(\bu')=b(\bu)}\exp\{nf(\hat{P}_{\bu'\bv})\}\ge
\exp\{n[f(\hat{P}_{\bu_0\bv})-\theta]\}\right\}.
\end{eqnarray}
Denote $\calC_n(\bu)=\{\bu':~b(\bu')=b(\bu)\}$, and for a given
conditional type $Q_{U'|V}$ of $\bu'$ given $\bv$, let
\begin{equation}
N(Q_{U'|V})=|\calC_n(\bu')\bigcap\calT(Q_{U'|V}|\bv)|.
\end{equation}
Obviously, $N(Q_{U'|V})$ is a binomial random variable with $|\calT(Q_{U'|V}|\bv)|\exe
e^{nH(U'|V)}$ trials and probability of success $e^{-nR}$.
Therefore,
\begin{eqnarray}
& &\mbox{Pr}\left\{\sum_{\bu':\bx(\bu')=\bx(\bu)}\exp\{nf(\hat{P}_{\bu'\bv})\}\ge
\exp\{n[f(\hat{P}_{\bu_0\bv})-\theta]\}\right\}\nonumber\\
&=&\mbox{Pr}\left\{\sum_{Q_{U'V}}N(Q_{U'V})e^{nf(Q_{U'V})}\ge
e^{n[f(\hP_{\bu_0\bv})-\theta]}\right\}\nonumber\\
&\exe&\max_{Q_{U'V}}\mbox{Pr}\left\{N(Q_{U'V})\ge
e^{n[f(\hP_{\bu_0\bv})-f(Q_{U'V})-\theta]}\right\}\nonumber\\
&=& \exp\{-nF_1(R,\hP_{\bu_0\bv},\theta)\}
\end{eqnarray}
where
\begin{eqnarray}
F_1(R,\hP_{\bu_0\bv},\theta)&=&\min_{Q_{U'V}}\{[R-H(U'|V)]_+:~f(\hP_{\bu_0\bv})-f(Q_{U'V})-\theta\le
[H(U'|V)-R]_+\}\nonumber\\
&=&\min_{Q_{U'V}}\{[R-H(U'|V)]_+:~\theta\ge f(\hP_{\bu_0\bv})-f(Q_{U'V})-
[H(U'|V)-R]_+\}
\end{eqnarray}
where the minimum over $\{Q_{U'V}\}$ is subject to the constraint that its
$V$--marginal coincides with $\hat{P}_{\bv}$.
Consequently, the contribution of $Z_1$ to the conditional probability of
error, which we
denote by $\bar{P}_{\mbox{\tiny e}1}(\bu_0,\bv,j_0,\bx_0,\by)$, is the
following:
\begin{eqnarray}
& &\bar{P}_{\mbox{\tiny e}1}(\bu_0,\bv,j_0,\bx_0,\by)\nonumber\\
&\exe&\int_0^\infty
e^{-n\theta}\exp\{-nF_1(R,\hP_{\bu_0\bv},\theta)\}\mbox{d}\theta\nonumber\\
&\exe&\int_{[f(\hP_{\bu_0\bv})-f(Q_{U'V})-[H(U'|V)-R]_+]_+}^\infty
e^{-n\theta}\exp\{-n[R-H(U'|V)]_+\}\mbox{d}\theta\nonumber\\
&\exe& e^{-nE_1(R,\hP_{\bu_0\bv})}
\end{eqnarray}
with
\begin{eqnarray}
E_1(R,\hP_{\bu_0\bv})&=&\min_{Q_{U'V}}\left\{[R-H(U'|V)]_++[f(\hP_{\bu_0\bv})-f(Q_{U'V})-
[H(U'|V)-R]_+]_+\right\}\nonumber\\
&=&\min_{Q_{U'V}}\left\{\begin{array}{ll}
[f(\hP_{\bu_0\bv})-f(Q_{U'V})+R-H(U'|V)]_+ & R < H(U'|V)\\
R-H(U'|V)+[f(\hP_{\bu_0\bv})-f(Q_{U'V})]_+ & R \ge H(U'|V)
\end{array}\right.\nonumber\\
&=&\min_{Q_{U'V}}[[f(\hP_{\bu_0\bv})-f(Q_{U'V})]_++R-H(U'|V)]_+
\end{eqnarray}
The overall contribution of $Z_1$ to the (unconditional) probability of error
is therefore
\begin{equation}
\bar{P}_{\mbox{\tiny e}1}\exe e^{-nE_2(R)}
\end{equation}
where
\begin{equation}
E_2(R)=\min_{Q_{UV}}\{D(Q_{UV}\|P_{UV})+E_1(R,Q_{UV})\},
\end{equation}
as defined also in Section 4.

We next move on to handle $Z_2$, which we have defined as
\begin{eqnarray}
Z_2&=&\sum_{\bu':~b(\bu')\ne b(\bu)}\exp\{n[f(\hat{P}_{\bu'\bv})+
g(\hat{P}_{\bx(\bu')\by})]\}\nonumber\\
&=&\sum_{\calT(Q_{U'|V}|\bv)}e^{nf(Q_{U'V})}\sum_{\calT(Q_{X'|Y}|\by)}e^{ng(Q_{X'Y})}
\sum_{\bu'\in\calT(Q_{U'|V}|\bv)}\calI[b(\bu')\ne
b(\bu)]\cdot\calI[\bX(\bu')\in \calT(Q_{X'|Y}|\by)]\nonumber\\
&\dfn&\sum_{\calT(Q_{U'|V}|\bv)}e^{nf(Q_{U'V})}\sum_{\calT(Q_{X'|Y}|\by)}e^{ng(Q_{X'Y})}
N(Q_{U'V},Q_{X'Y}).
\end{eqnarray}
Now, for a given channel code $\calC_n$,
$N(Q_{U'V},Q_{X'Y})$ is a binomial random variable with exponentially $e^{nH(U'|V)}$
trials and probability of success
$(1-e^{-nR})|(\calC_n\setminus\{\bx_0\})
\cap\calT(\bx'|\by)|/(|\calC_n|-1)\exe e^{-nR}|(\calC_n\setminus\{\bx_0\})\cap\calT(\bx'|\by)|\dfn
e^{-nS(Q_{X'Y})}$.
Thus,
\begin{eqnarray}
& &\mbox{Pr}\left\{\sum_{\calT(Q_{U'|V}|\bv)}e^{nf(Q_{U'V})}
\sum_{\calT(Q_{X'|Y}|\by)}e^{ng(Q_{X'Y})}
N(Q_{U'V},Q_{X'Y})\ge
\exp\{n[f(\hP_{\bu_0\bv})+g(\hP_{\bx_0\by})-\theta]\}\right\}\nonumber\\
&\exe&\max_{Q_{U'V},Q_{X'Y}}\mbox{Pr}\left\{N(Q_{U'V},Q_{X'Y})\ge
\exp\{n[f(\hP_{\bu_0\bv})+g(\hP_{\bx_0\by})-f(Q_{U'V})-g(Q_{X'Y})-\theta]\}\right\}.
\end{eqnarray}
We henceforth use the shorthand notation
$h(Q_{U'V},Q_{X'Y})=f(Q_{U'V})+g(Q_{X'Y})$, as defined in Section 4.
The exponential order of the last probability is given by
\begin{eqnarray}
F_2(Q_{UV},Q_{XY},\theta)&=&
\min_{Q_{U'V},Q_{X'Y}}\left\{[S(Q_{X'Y})-H(U'|V)]_+:~\theta\ge
h(Q_{UV},Q_{XY})-\right.\nonumber\\
& &\left.h(Q_{U'V},Q_{X'Y})-[H(U'|V)-S(Q_{X'Y})]_+\right\}
\end{eqnarray}
Now,
\begin{eqnarray}
& &\max_{Q_{U'V},Q_{X'Y}}\int_0^\infty
e^{-n\theta}\exp\{-nF_2(Q_{UV},Q_{XY},\theta)\}\mbox{d}\theta\nonumber\\
&\exe&\max_{Q_{U'V},Q_{X'Y}}\int_{[h(Q_{UV},(Q_{XY})-
h(Q_{U'V},(Q_{X'Y})-[H(U'|V)-S(Q_{X'Y})]_+]_+}^\infty
e^{-n\theta}\exp\{-n[S(Q_{X'Y})-H(U'|V)]_+\}\mbox{d}\theta\nonumber\\
&\exe&\exp\{-n\min_{Q_{U'V},Q_{X'Y}}E_3(Q_{UV},Q_{XY},Q_{U'V},Q_{X'Y},S(Q_{X'Y}))\}
\end{eqnarray}
where for a given $S$,
\begin{eqnarray}
& &E_3(Q_{UV},Q_{XY},Q_{U'V},Q_{X'Y},S)\nonumber\\
&\dfn&
\left\{[S-H(U'|V)]_++[h(Q_{UV},Q_{XY})-
h(Q_{U'V},Q_{X'Y})-[H(U'|V)-S]_+]_+\right\}\nonumber\\
&=&\min_{Q_{U'V},Q_{X'Y}}\left\{\begin{array}{ll}
[h(Q_{UV},Q_{XY})-h(Q_{U'V},Q_{X'Y})+S-H(U'|V)]_+ &
S< H(U'|V)\\
S-H(U'|V)+[h(Q_{UV},Q_{XY})-h(Q_{U'V},Q_{X'Y})]_+ &
S\ge H(U'|V)
\end{array}\right.\nonumber\\
&=&[[h(Q_{UV},Q_{XY})-h(Q_{U'V},Q_{X'Y})]_++S-H(U'|V)]_+
\end{eqnarray}
It remains to average this expression w.r.t.\ the randomness of the channel
code $\calC_n$.
For a given $Q_{X'Y}$, it follows from the definition of $S(Q_{X'Y})$
that $S(Q_{X'Y})=R-\frac{1}{n}\log N(Q_{X'Y})$ where $N(Q_{X'Y})$ is the
number of codewords in $\calC_n\setminus\{\bx_0\}$ whose joint type with $\by$ is $Q_{X'Y}$.
Now, $N(Q_{X'Y})$ is a binomial random variable with $e^{nR}$ trials and probability of
success of the exponential order of $e^{-nI(X';Y)}$. Therefore, the expectation of
$\exp\{-nE_3(Q_{UV},Q_{XY},Q_{U'V},Q_{X'Y},S(Q_{X'Y}))\}$ w.r.t.\ the
randomness of the code is assessed as follows. Let $\epsilon > 0$ be
arbitrarily small. Then the desired average is upper bounded by
\begin{eqnarray}
& &\sum_{i\ge 0}\mbox{Pr}\left\{e^{ni\epsilon}\le N(Q_{X'Y})<
e^{n(i+1)\epsilon}\right\}\cdot
\exp\{-nE_3(Q_{UV},Q_{XY},Q_{U'V},Q_{X'Y},R-(i+1)\epsilon)\}\nonumber\\
&\exe& \max_{0\le i\le [R-I(X';Y)]_+/\epsilon} \exp\{-n[I(X';Y)-R]_+\}\cdot
\exp\{-nE_3(Q_{UV},Q_{XY},Q_{U'V},Q_{X'Y},R-(i+1)\epsilon)\}\nonumber\\
&\exe&\exp\{-n[I(X';Y)-R]_+\}\cdot
\exp\{-nE_3(Q_{UV},Q_{XY},Q_{U'V},Q_{X'Y},R-[R-I(X';Y)]_+)-\epsilon\}\nonumber\\
&=&\exp\{-n
E_3(Q_{UV},Q_{XY},Q_{U'V},Q_{X'Y},I(X';Y)-\epsilon)\},
\end{eqnarray}
but since $\epsilon > 0$ is arbitrary,
$E_3(Q_{UV},Q_{XY},Q_{U'V},Q_{X'Y},I(X';Y))$ can be approached as closely as
desired. We henceforth omit the term $\epsilon$ in $E_3$ and denote
\begin{equation}
E_4(Q_{UV},Q_{XY})=\min_{Q_{U'V},Q_{X'Y}}
E_3(Q_{UV},Q_{XY},Q_{U'V},Q_{X'Y},I(X';Y)),
\end{equation}
the overall contribution of $Z_2$ to the average probability of error is
of the exponential order of $e^{-nE_5}$, where
\begin{equation}
E_5=\min_{Q_{UV},Q_{XY}}[D(Q_{UV}\|P_{UV})+D(Q_{Y|X}\|W|P)+E_4(Q_{UV},Q_{XY})].
\end{equation}
Finally, the overall exponent is
\begin{equation}
E(R)=\min\{E_2(R),E_5\},
\end{equation}
which completes the proof of Theorem 1.

\section*{Appendix B}
\renewcommand{\theequation}{B.\arabic{equation}}
    \setcounter{equation}{0}

{\it Proof of Eq.\ (\ref{appb}).}

We show that the vast majority of codes have
$Z_m(\by)\ge \exp\{n\alpha(R-\epsilon,\hP_{\by})\}$ for all $m$ and $\by$.
First, observe that
\begin{equation}
Z_m(\by)=\sum_{m'\ne m}\exp\{ng(\hP_{\bx_{m'}\by})\}=\sum_Q
N_{\by}(Q)e^{ng(Q)}.
\end{equation}
Thus, considering the randomness of $\{\bX_{m'}\}$,
\begin{eqnarray}
& &\mbox{Pr}\left\{Z_m(\by)\le
\exp\{n\alpha(R-\epsilon,\hP_{\by})\}\right\}\nonumber\\
&=&\mbox{Pr}\left\{\sum_Q N_{\by}(Q)e^{ng(Q)}\le
\exp\{n\alpha(R-\epsilon,\hP_{\by})\}\right\}\nonumber\\
&\le&\mbox{Pr}\left\{\max_Q N_{\by}(Q)e^{ng(Q)}\le
\exp\{n\alpha(R-\epsilon,\hP_{\by})\}\right\}\nonumber\\
&=&\mbox{Pr}\bigcap_Q\left\{N_{\by}(Q)e^{ng(Q)}\le
\exp\{n\alpha(R-\epsilon,\hP_{\by})\}\right\}\nonumber\\
&=&\mbox{Pr}\bigcap_Q\left\{N_{\by}(Q)\le
\exp\{n[\alpha(R-\epsilon,\hP_{\by})-g(Q)]\}\right\}.
\end{eqnarray}
Now, $N_{\by}(Q)$ is a binomial random variable with $e^{nR}$ trials and success rate of
the exponential order of $e^{-nI(Q)}$. 
We now argue that
by the very definition of $\alpha(R-\epsilon,\hP_{\by})$, there
must exist
some $Q_{X|Y}^*$ such that for $Q^*=\hP_{\by}\times Q_{X|Y}^*$,
$I(Q^*) \le R-\epsilon$ and $R-\epsilon-I(Q^*) \ge
\alpha(R-\epsilon,\hP_{\by})-g(Q^*)$. To see why this is true, assume
conversely,
that for every $Q_{X|Y}$, which defines $Q=\hP_{\by}\times Q_{X|Y}$, 
either $I(Q) > R-\epsilon$ or $R-I(Q)-\epsilon <
\alpha(R-\epsilon,\hP_{\by})-g(Q)$, which means that for every $Q$
\begin{equation}
R-\epsilon<\max\{I(Q),I(Q)+\alpha(R-\epsilon,\hQ_{\by})-g(Q)\}=I(Q)+
[\alpha(R-\epsilon,\hQ_{\by})-g(Q)]_+
\end{equation}
which implies in turn that for every $Q_{X|Y}$ there exists $t\in[0,1]$ such that
\begin{equation}
R-\epsilon<I(Q)+t[\alpha(R-\epsilon,\hQ_{\by})-g(Q)]
\end{equation}
or equivalently,
\begin{eqnarray}
\alpha(R-\epsilon,\hP_{\by})&>&\max_{Q_{X|Y}}\min_{0\le t\le
1}g(Q)+\frac{R-I(Q)-\epsilon}{t}\nonumber\\
&=&\max_{Q_{X|Y}}\left\{\begin{array}{ll}
g(Q)+R-I(Q)-\epsilon & I(Q)\le R-\epsilon\\
-\infty & I(Q) > R-\epsilon\end{array}\right.\nonumber\\
&=&\max_{\{Q_{X|Y}:~I(Q)\le R-\epsilon\}}[g(Q)-I(Q)]+R-\epsilon\nonumber\\
&\equiv& \alpha(R-\epsilon,\hP_{\by}),
\end{eqnarray}
which is a contradiction. 
Let then $Q_{X|Y}^*$ be as defined above. Then,
\begin{eqnarray}
& &\mbox{Pr}\bigcap_Q\left\{N_{\by}(Q)
\le \exp\{n[\alpha(R-\epsilon,\hP_{\by})-g(Q)]\}\right\}\nonumber\\
&\le&
\mbox{Pr}\left\{N_{\by}(Q^*)\le
\exp\{n[\alpha(R-\epsilon,\hP_{\by})-g(Q^*)]\}\right\}.
\end{eqnarray}
Now, we know that $I(Q^*) \le R-\epsilon$ and $R-I(Q^*)-\epsilon \ge
\alpha(R-\epsilon,\hP_{\by})-g(Q^*)$.
By the Chernoff bound,
the probability in question is upper bounded by
\begin{equation}
\exp\left\{-e^{nR}D(e^{-an}\|e^{-bn})\right\},
\end{equation}
where $a=R+g(Q^*)-\alpha(R-\epsilon,\hP_{\by})$ and
$b=I(Q^*)$. Noting that $a-b\ge\epsilon$,
we can easily lower bound the binary divergence as follows (see \cite[Section
6.3]{fnt}):
\begin{eqnarray}
D(e^{-an}\|e^{-bn})&\ge&e^{-bn}\{1-e^{-(a-b)n}[1+n(a-b)]\}\nonumber\\
&\ge&e^{-nI(Q^*)}[1-e^{-n\epsilon}(1+n\epsilon)],
\end{eqnarray}
where in the last passage, we have used the decreasing monotonicity of the
function $f(t)=(1+t)e^{-t}$ for $t\ge 0$.
Thus,
\begin{eqnarray}
\mbox{Pr}\left\{N_{\by}(Q^*)\le
\exp\{n[\alpha(R,\hP_{\by})-g(Q^*)-\epsilon]\}\right\}&\le&
\exp\left\{-e^{nR}\cdot
e^{-nI(Q)}[1-e^{-n\epsilon}(1+n\epsilon)]\right\}\nonumber\\
&\le&\exp\left\{-e^{n\epsilon}[1-e^{-n\epsilon}(1+n\epsilon)]\right\}\nonumber\\
&=&\exp\left\{-e^{n\epsilon}+n\epsilon+1\right\},
\end{eqnarray}
which completes the proof of eq.\ (\ref{appb}).

\section*{Appendix C}
\renewcommand{\theequation}{C.\arabic{equation}}
    \setcounter{equation}{0}

\noindent
{\it Proof of Eq.\ (\ref{appc}).}
The proof is very similar to the proof of Theorem 2 in \cite{list}, but there
is a small twist due to the limitation to codes in $\calG_\epsilon$ herein. 
Consider the uniform random selection of codebooks $\{\calC_n\}$ in
$\calG_\epsilon$.
For every given code $\calC_n\in\calG_\epsilon$ and a given message $m$,
let $N_m(Q_{XX'},\calC_n)$ be the number of
$\{m'\}$, all different from $m$,
for which $(\bx_m,\bx_{m'})$ has a given joint empirical
distribution $Q_{XX'}$ of a pair of random variables taking on
values in $\calX^2$, whose single--letter marginals
(which are the individual empirical distributions of the various codewords)
all coincide with $Q_X$.
Our goal here is to show that
for every $\epsilon > 0$ and sufficiently large $n$,
there exists a code $\calC_n\in\calG_\epsilon$ of rate (essentially) $R$,
that satisfies, for every message $m$ and every $Q_{XX'}$,
eq.\ (\ref{appc}), namely,
\begin{eqnarray}
N_m(Q_{XX'},\calC_n)&\le&N^*(Q_{XX'})\nonumber\\
&\dfn&\left\{\begin{array}{ll}
\exp\{n[R-I_Q(X;X')+\epsilon]\} & R\ge
I_Q(X;X')-\epsilon\\
0 & R< I_Q(X;X')-\epsilon\end{array}\right.
\end{eqnarray}
To see why this is true, consider a random selection of the code $\calC_n$
within $\calG_\epsilon$.
Then, obviously,
\begin{eqnarray}
\overline{N(Q_{XX'})}&\dfn&
\frac{1}{M}\sum_{m=0}^{M-1}\bE\{N_m(Q_{XX'},\calC_n)|\calG_\epsilon\}\\
&\le&\frac{1}{M}\sum_{m=0}^{M-1}\frac{\bE\{N_m(Q_{XX'},\calC_n)\}}{\mbox{Pr}\{\calG_\epsilon\}}\\
&\exe&\bE\{N_0(Q_{XX'},\calC_n)\}\\
&=&
M\cdot\mbox{Pr}\{(\bX,\bX')\in\calT(Q_{XX'})\}\\
&=& M\cdot\frac{|\calT(Q_{XX'})|}{|\calT(Q_X)|^2}\\
&\exe& M\cdot\frac{\exp\{nH_Q(X,X')\}}{e^{2nH_Q(X)}}\\
&=&\exp\{n[R-I_Q(X;X')]\},
\end{eqnarray}
where the unconditional expectation in the second and third lines corresponds
to uniform random selection across the whole class of fixed composition codes,
not just to $\calG_\epsilon$.
It follows then that in the ensemble of all randomly selected codes within
$\calG_\epsilon$:
\begin{eqnarray}
& &\mbox{Pr}\bigcup_{Q_{XX'}}\left\{\calC_n:~\frac{1}{M}\sum_{m=0}^{M-1}N_m(Q_{XX'},
\calC_n)>
\exp\{n[R-I_Q(X;X')+\epsilon/2]\}\bigg|\calC_n\in\calG_\epsilon\right\}\nonumber\\
&\le&\sum_{Q_{XX'}}\mbox{Pr}\left\{\calC_n:~\frac{1}{M}
\sum_{m=0}^{M-1}N_m(Q_{XX'},
\calC_n)>
\exp\{n[R-I_Q(X;X')+\epsilon/2]\}\bigg|\calC_n\in\calG_\epsilon\right\}\nonumber\\
&\lexe&\sum_{Q_{XX'}}\frac{\overline{N(Q_{XX'})}}
{\exp\{n[R-I_Q(X;X')+\epsilon/2]\}}\nonumber\\
&\lexe&\sum_{Q_{XX'}}e^{-n\epsilon/2}\nonumber\\
&\le&(n+1)^{|\calX|^2}\cdot e^{-n\epsilon/2}\to 0,
\end{eqnarray}
which means that there exists a code in $\calG_\epsilon$ (and in fact, for
almost every such code),
\begin{equation}
\frac{1}{M}\sum_{m=0}^{M-1}N_m(Q_{XX'},
\calC_n)\le \exp\{n[R-I_Q(X;X')+\epsilon/2]\}~~~\forall
Q_{XX'}.
\end{equation}
For a given such code and every given $Q_{XX'}$, there must then
exist
at least $(1-e^{-n\epsilon/2})\cdot M$ values of $m$ such that
\begin{equation}
N_m(Q_{XX'},
\calC_n)\le \exp\{n[R-I_Q(X;X')+\epsilon]\}.
\end{equation}
Upon eliminating the exceptional codewords from the code, for all
$Q_{XX'}$, one ends up with at least
$[1-(n+1)^{|\calX|^2}e^{-n\epsilon/2}]\cdot M$ for which
\begin{equation}
N_m(Q_{XX'},
\calC_n)\le \exp\{n[R-I_Q(X;X')+\epsilon]\}~~~~\forall
Q_{XX'}.
\end{equation}
Let $\calC_n'$ denote the sub--code formed by these
$[1-(n+1)^{|\calX|^2}e^{-n\epsilon/2}]\cdot M$
remaining codewords. Since $N_m(Q_{XX'},
\calC_n')\le N_m(Q_{XX'},
\calC_n)$, then the sub--code $\calC_n'$ certainly satisfies
\begin{equation}
N_m(Q_{XX'},
\calC_n')\le \exp\{n[R-I_Q(X;X')+\epsilon]\}~~~~\forall m,
Q_{XX'}.
\end{equation}
Finally, observe that since $N_m(Q_{XX'},
\calC_n')$ is a non--negative integer, then for $Q_{XX'}$ with
$R-I_Q(X;X')+\epsilon< 0$, the last inequality means
$N_m(Q_{XX'},
\calC_n')=0$, in which case the r.h.s.\ of the last equation becomes
$N^*(Q_{XX'})$. Thus, we have shown that there exists a code
$\calC_n'$ of
size $M'=[1-(n+1)^{|\calX|^2}e^{-n\epsilon/2}]\cdot e^{nR}$ for which all
codewords satisfy $N_m(Q_{XX'},
\calC_n')\le N^*(Q_{XX'})$ for all joint types $Q_{XX'}$.


\clearpage
\begin{thebibliography}{AA}

\bibitem{Csiszar80}
I.~Csisz\'ar, ``Joint source--channel error exponent,'' {\it Problems of
Control and Information Theory}, vol.\ 9, no.\ 5, pp.\ 315--328, 1980.

\bibitem{CK11}
I.~Csisz\'ar and J.~K\"orner, {\it Information Theory: Coding Theorems for
Discrete Memoryless Systems}, Second Edition, Cambridge University Press,
2011.

\bibitem{CKM77}
I.~Csisz\'ar, J.~K\"orner, and K.~Marton, ``A new look at the error exponent
of a discrete memoryless channel,'' {\it Proc.\ ISIT `77}, p.\ 107 (abstract),
Cornell University, Itacha, New York, U.S.A., 1977.

\bibitem{CN95}
I.~Csisz\'ar and P.~Narayan, ``Channel capacity for a given decoding 
metric,'' {\it IEEE Trans.\ Inform.\ Theory}, vol.\ 45, no.\ 1, pp.\ 35--43,
January 1995.

\bibitem{Gallager68}
R.~G.~Gallager, {\it Information Theory
and Reliable Communication}, New York, Wiley 1968.

\bibitem{Hui83}
J.~Hui, {\it Fundamental issues of multiple accessing}, Ph.D.\ dissertation,
MIT, 1983.

\bibitem{fnt}
N.~Merhav, ``Statistical physics and information theory,''
{\it Foundations and Trends in
Communications and Information Theory}, vol.\ 6, nos.\ 1--2, pp.\ 1--212,
2009.

\bibitem{REM1st}
N.~Merhav, ``Relations between random coding exponents and
the statistical physics of random codes,''
{\it IEEE Trans.\ Inform.\ Theory}, vol.\ 55, no.\ 1, pp.\ 83--92, January
2009.

\bibitem{swlist}
N.~Merhav, ``Erasure/list exponents for Slepian--Wolf decoding,'' {\it IEEE
Trans.\ Inform.\ Theory}, vol.\ 60, no.\ 8, pp.\ 4463--4471, August 2014.

\bibitem{bid}
N.~Merhav, ``Exact random coding exponents of optimal bin index decoding,''
{\it IEEE Trans.\ Inform.\ Theory}, vol.\ 60, no.\ 10, pp.\ 6024--6031,
October 2014.

\bibitem{jscuniv}
N.~Merhav, ``Universal decoding for source--channel coding with side
information,'' submitted to {\it IEEE Trans.\ Inform.\ Theory}, July 2015.
Available on--line at {\tt http://arxiv.org/pdf/1505.01255.pdf} 

\bibitem{list}
N.~Merhav, ``List decoding -- random coding exponents
and expurgated exponents,''
{\it IEEE Trans.\ Inform.\ Theory},
vol.\ 60, no.\ 11, pp.\ 6749--6759, November 2014.

\bibitem{MKLS94}
N.\ Merhav, G.\ Kaplan, A.\ Lapidoth, and S.\ Shamai (Shitz),
``On information rates for mismatched decoders,''
{\em IEEE Trans. Inform. Theory},
vol.\ 40, no.\ 6, pp.\ 1953-1967, November 1994.

\bibitem{Rujan93}
P.~Ruj\'an, ``Finite temperature error--correcting codes,'' 
{\it Phys.\ Rev.\ Let.}, vol.\ 70, no.\ 19, pp.\ 2968--2971, May 1993.

\bibitem{SMF15}
J.~Scarlett, A.~Martin\'ez and A.~G.~i F\'abregas, ``The likelihood decoder:
error exponents and mismatch,'' {\it Proc.\ 2015 IEEE International Symposium
on Information Theory (ISIT 2015)}, pp.\ 86--90, Hong Kong, June 2015.

\bibitem{SCP14}
E.~C.~Song, P.~Cuff and H.~V.~Poor, ``The likelihood encoder for lossy
compression,'' [{\tt http://arxiv.org/abs/1408.4522}], 2014.

\bibitem{YAG13}
M.~H.~Yassaee, M.~R.~Aref and A.~Gohari, ``A technique for deriving one--shot
achievability results in network information theory,'' [{\tt
http://arxiv.org/abs/1303.0696}], 2013.

\end{thebibliography}
\end{document}